\documentclass[12pt]{iopart}

\usepackage{iopams}
\usepackage{braket}
\usepackage{amsfonts}
\usepackage{color}
\usepackage[dvipsnames]{xcolor}
\usepackage{amssymb}
\usepackage{mathrsfs}
\usepackage{graphicx}
\usepackage{lmodern}
\usepackage{lineno}
\usepackage{hyperref}
\usepackage[normalem]{ulem}
\usepackage{soul}

\newcommand{\beq}{\begin{eqnarray}}
\newcommand{\eeq}{\end{eqnarray}}

\def\keywords#1{\vspace{10pt}
     \begin{indented}
     \item[]\rm Keywords: #1\par
     \end{indented}}

\begin{document}

%\linenumbers

%\draftstring{DRAFTv4, 15/12/16}
%\draftfontsize{60}
%%\draftfontfamily{hlh}
%\draftfontfamily{ptm}
%\draftangle{45}
%\definecolor{mycolor}{rgb}{.855,.855,1}
%\draftcolor{mycolor}
%\draftfontattrib{\upshape}

\title{Quantum energy teleportation in phase space quantum mechanics}
\author{MM S\'anchez-C\'ordova$^{1}$ and Jasel Berra-Montiel$^{1,2}$}

\address{$^{1}$ Facultad de Ciencias, Universidad Aut\'onoma de San Luis 
Potos\'{\i} \\
Campus Pedregal, Av. Parque Chapultepec 1610, Col. Privadas del Pedregal, San
Luis Potos\'{\i}, SLP, 78217, Mexico}
\address{$^2$ Dipartimento di Fisica ``Ettore Pancini", Universit\'a degli studi di Napoli ``Federico II", Complesso Univ. Monte S. Angelo, I-80126 Napoli, Italy}

\eads{\mailto{\textcolor{blue}{maria.cordova@uaslp.mx}},\ 
\mailto{\textcolor{blue}{jasel.berra@uaslp.mx}}\
}

% \received{20 October 2016}
% \vspace{-2ex}
% \accepted{}
% \vspace{-2ex}
% \published{}

\begin{abstract}
In this paper, we investigate the Quantum Energy Teleportation protocol within the phase space formulation of quantum mechanics. By employing the Wigner quasi-probability distribution and the star product, we show that the teleported energy is proportional to the amount of entanglement present in the initial ground state. Further, we introduce the Husimi $Q$-function on a Bloch coherent state in order to determine the Wehrl entropy of the system. Finally, the Wherl entropy enable us to compute the consumption of coherence and entanglement throughout the protocol.  
\end{abstract}

%\ams{81S30, 46F10, 53D55, 81S40}
%\pacs{}
%\ams{81S30, 53D55, 70H45}

% \maketitle

\section{Introduction}
The phase space representation of quantum mechanics, also referred as deformation quantization by many authors, provides a general method to investigate quantum systems implementing notions from classical physics \cite{Wigner},\cite{Weyl}. The distinctive characteristic of this quantization approach lies in the role assumed by the algebra of quantum observables, which is not represented by a set of operators defined on a Hilbert space \cite{Moyal}. Instead the observables are associated with smooth complex-valued functions on the classical phase space, where the usual commutative point-wise product is replaced by a non-commutative product termed as the star-product \cite{Bayen}. A central feature within this formulation is determined by the Wigner distribution function. This function provides a phase space representation of the density operator and captures all auto-correlation properties and transition amplitudes of a quantum system. The phase space formalism has undoubtedly made significant contributions not only to pure mathematics \cite{Kontsevich}, \cite{Waldmann}, but it has also proved to be a reliable technique in the understanding of many physical complex systems \cite{Zachos},\cite{Cahill}, including more recently, finite dimensional multi-qubit systems \cite{Tilma},\cite{Rundle}, distinct aspects of the loop representation of quantum cosmology and quantum gravity \cite{Berra},\cite{Berra2},\cite{Berra3},\cite{BerraW} and novel connections with the path integral formalism \cite{Berra4}, \cite{Berra5}, to cite a few examples. However, despite the broad relevance of the phase space formalism, a limited understanding remains regarding its applicability to recent discovered phenomena involving the interplay between quantum information theory and quantum field theory, such as Quantum Energy Teleportation (QET) \cite{Hotta0},\cite{Hotta3}, black hole loss information problem \cite{Preskill} and entanglement harvesting \cite{Salton}, among others.

Following some ideas formulated in \cite{Tilma},\cite{Brif}, in this paper we propose to analyse the protocol of Quantum Energy Teleportation within the phase space formalism of quantum mechanics. This protocol, originally developed within the framework of quantum field theory and many body systems, were first propose by Hotta \cite{Hotta0} as a method to transfer energy, through local operations and classical communication (LOCC) with an entangled partner, by locally generating regions with negative energy density. This technique, based on the entanglement of the ground state, has recently found many applications in cold trapped ions \cite{cold}, relativistic quantum information \cite{MartinEn} and black hole evaporation \cite{HottaBH}. With the aim to describe the QET protocol, we compute the variation of its associated Wigner quasi-probability distribution under the sequence of local unitarities applied on the initial ground state. As we will demonstrate below, the Wigner and the Husimi phase space functions enable us to quantify the amount of coherence and entanglement consumed throughout the protocol, with the aim to extract energy without requiring any physical transport between the source and the receiver. Finally, a correlation between the entanglement negativity and the Wehrl entropy for the system is discussed.

The paper is organized as follows, in section \ref{sec:WF} we briefly introduce the fundamental concepts of phase space quantization, Wigner functions and star products in multi-qubit systmes. Then, in section \ref{sec:QET} we analyse the QET protocol within the phase space formalism, focusing our attention on the evolution of the entanglement. Finally, we introduce some concluding remarks in section \ref{sec:conclu}.                 	 

\section{Phase space quantization and Wigner functions}
\label{sec:WF}

The phase space formulation of quantum mechanics possesses an extensive history. In 1932 Wigner introduced his renowned function, which has been widely applied in numerous fields of physics, mathematics, electronics and geophysics \cite{Wigner},\cite{Weyl}. In 1949, Moyal found that the inverse of the Weyl correspondence rule can be obtained by using the Wigner transform, enabling the association of an operator defined on a Hilbert space with a function on the classical phase space \cite{Moyal}. Consequently, the expectation value of an operator can be expressed as the statistical average of the corresponding phase-space function, where the probability density is determined by the Wigner function related to the density matrix of the quantum state. In this manner, quantum mechanics can be formally framed as a statistical-like theory on classical phase space. This formalism, known as phase space quantum mechanics proves to be equivalent to the standard Schr\"odinger and Heisenberg representations. Lately, in the seminal papers \cite{Bayen}, it was shown that the phase space formulation of quantum mechanics inherently possesses a noncommutative deformation of the algebraic structure of classical observables, where the usual pointwise product of functions should be replaced by the so called star product (also known as the Moyal product) in order to recover the operator algebra structure of quantum mechanics. This approach currently referred to as deformation quantization consists in a entirely new formulation of quantum mechanics in terms of phase space algebras instead of employing operators on Hilbert spaces. Most recently, M. Kontsevich proved that there is a one-to-one correspondence between equivalent classes  of star products and Poisson structures, showing that every Poisson manifold, and as a consequence every classical system, admits a canonical quantization, which is in fact unique up to formal equivalence \cite{Kontsevich}.

Originally, the conventional expression for the Wigner function was restricted to continuous non-relativistic	 spinless systems, such as quantum particles on flat phase spaces. It is customarily introduced via the Weyl-Wigner transform, which maps an operator $\hat{A}$ defined on a Hilbert space to a classical phase space function $W_{\hat{A}}(x,p)$ \cite{Weyl}, as
               
\begin{equation}\label{WeylA}
	W_{\hat{A}}(x,p)=\int_{\mathbb{R}}dy\,\e^{-\frac{i}{\hbar}yp}\braket{x-\frac{y}{2}|\hat{A}|x-\frac{y}{2}}. 
\end{equation}
For simplicity, we will focus on systems with a single degree of freedom, but the generalization to more dimensions follows straightforwardly. Now, let $\hat{\rho}$ be a normalized density operator associated with the quantum state $\psi(x)$, that is, a self-adjoint  positive semi-definite operator with trace one. By making use of the  Weyl-Wigner transform (\ref{WeylA}), one may see that its corresponding phase space function up to multiplication by a constant reads
\begin{equation}
W_{\hat{\rho}}(x,p)=\frac{1}{2\pi\hbar}\int_{\mathbb{R}}dy\,e^{-\frac{i}{\hbar}yp}\psi^{*}\left(x-\frac{y}{2}\right) \psi\left( x+\frac{y}{2} \right). 
\end{equation}
This special representation of the density operator embodies the celebrated Wigner function \cite{Wigner}, which allows us to characterize both pure and mixed quantum states by using probability distributions in phase space. However, in order to reflect the probabilistic nature of quantum theory, the Wigner distribution may take negative values in certain regions of phase space. As a result, it cannot be interpreted as a true probability density and is often referred to as a quasi-probability distribution in the literature \cite{Zachos}. Furthermore, the Wigner function plays an essential role in determining the expectation values of operators. This is achieved by integrating the associated functions over the phase space as follows
\begin{equation}\label{expectation}
	\braket{\psi,\hat{A}\psi}=\int_{\mathbb{R}^{2}}dx\,dp\,W_{\hat{\rho}}(x,p)W_{\hat{A}}(x,p).
\end{equation}
Moreover, the Wigner function obeys the normalization condition
\begin{equation}
\int_{\mathbb{R}^{2}}dx\,dp\,W_{\hat{\rho}}(x,p)=\tr \hat{\rho}=1.
\end{equation}
These expressions underscore the utility of the Wigner function in connecting quantum mechanical operators with their classical counterparts, enabling a bridge between both classical and quantum descriptions.
This formalism can also be written in terms of the displacement and parity operators given by the relation
\begin{equation}
W_{\hat{\rho}}(x,p)=\frac{1}{\pi\hbar}\tr\left[\hat{\rho}\hat{D}(x,p)\hat{\Pi}\hat{D}^{\dagger}(x,p)\right], 
\end{equation}
where the operator $\hat{D}(x,p)$ denotes the unitary displacement operator and $\hat{\Pi}$ represents the parity operator which acts according to
\begin{equation}
\hat{\Pi}\ket{x}=-\ket{x},\;\;\;\hat{\Pi}\ket{p}=-\ket{p},
\end{equation}  
where $\ket{x}$ and $\ket{p}$ comprise a complete set of eigenstates of the position and momentum operators respectively \cite{Agarwal}. 
It has been recently understood that a generalization of the phase space quantization approach for quantum systems with intrinsic symmetry groups, such as spin systems, can be achieved by using the Stratonovich-Weyl correspondence \cite{Brif}. The premise behind the Stratonovich-Weyl correspondence implies that a linear bijective mapping between operators acting on a Hilbert space and functions defined on phase space, also called symbols, can be realized via an appropriate kernel $\hat{\Delta}(\Omega)$, where $\Omega$ is any parametrization of the phase space, which satisfies the following physical motivated postulates \cite{Stratonovich}:  
\begin{enumerate}
\item The mapping $W_{\hat{A}}(\Omega)=\tr\left[ \hat{A}\hat{\Delta}(\Omega)\right] $ is a one-to-one linear map.
\item $W_{\hat{A}}(\Omega)$ is a real valued function, which implies that the operator kernel $\hat{\Delta}(\Omega)$ must be hermitian.
\item $W_{\hat{A}}(\Omega)$ satisfies standarization, which means that the integral over all phase space $\int_{\Omega}W_{\hat{A}}(\Omega)d\Omega=\tr \hat{A}$ exists and $\int_{\Omega}\hat{\Delta}(\Omega)d\Omega=\hat{1}$.
\item Traciality, that is $\int_{\Omega}W_{\hat{A}}(\Omega)W_{\hat{B}}(\Omega)d\Omega=\tr\left[\hat{A}\hat{B} \right] $.
\item Covariance, which indicates that if an operator $\hat{A}$ is invariant under global unitary operators, then $W_{\hat{A}}(\Omega)$ also remains invariant. 
\end{enumerate}  
In the case of systems with continuous variables, the kernel operator $\hat{\Delta}(\Omega)$ proves to be proportional to $\hat{D}(x,p)\hat{\Pi}\hat{D}^{\dagger}(x,p)$, where the parametrization is given by $\Omega=\left\lbrace x,p\right\rbrace$. For other systems, it is crucial to select both the kernel operator and the parametrization phase space coordinates in a manner that the symmetries of the physical system under study are precisely captured \cite{Tilma}.

\noindent For the case of a two level quantum system, such as a qubit over the Bloch sphere with continuous degrees of freedom, the corresponding parity operator reads
\begin{equation}\label{Parity}
	\hat{\Pi}_q=\hat{1}_{2}+\sqrt{3}\hat{\sigma}_z, 
\end{equation}
where $\hat{1}_{2}$ denotes the $2\times 2$ identity matrix and $\hat{\sigma}_{z}$ stands for the Pauli z operator. This form of the spin parity operator arises from analysing the $SU(N)$ coherent states in the complex projective space \cite{Nemoto}, \cite{Perelomov}. On the other hand, the analogous of the displacement operator for spin systems is given by the $SU(2)$ rotation operator  
\begin{equation}\label{SU2Rotation}
	\hat{U}(\phi,\theta,\Phi)=exp(-i\hat{\sigma}_z\frac{\phi}{2})exp(-i\hat{\sigma}_y \frac{\theta}{2})exp(-i\hat{\sigma}_z\frac{\Phi}{2}), 
\end{equation}
where $0\leq\theta\leq \pi$, $0\leq\phi\leq 2\pi$ and $0\leq\Phi\leq 4\pi$ are the conventional Euler angles \cite{Sakurai}. Then, following the Stratonovich-Weyl postulates the corresponding kernel for a qubit system follows   
\begin{eqnarray}\label{Delta}
	\hat{\Delta}_q(\theta,\phi)&=&\frac{1}{2}\hat{U}(\phi,\theta,\Phi)\hat{\Pi}_q\hat{U}(\phi,\theta,\Phi)^{\dagger}, \nonumber\\
&=&\frac{1}{2}\left(
\begin{array}{cc}
	1+\sqrt{3}\cos{\theta} & \sqrt{3}e^{i\phi}\sin{\theta} \\
	\sqrt{3}e^{-i\phi}\sin{\theta} & 1-\sqrt{3}\cos{\theta} 
\end{array}
\right).		 
\end{eqnarray}
It is noteworthy that the kernel $\hat{\Delta}_q(\theta,\phi)$ does not depend explicitly of the parameter $\Phi$, this allow us to restrict the phase space functions on a 2-sphere $\mathbb{S}^{2}$,  which turns out to be equivalent to the representation used for the Bloch sphere \cite{Tilma}, however, for larger Hilbert spaces the $\Phi$ degree of freedom must be considered \cite{Overview}. Subsequently, by means of the kernel $\hat{\Delta}_{q}(\theta,\phi)$ we can obtain the symbol associated to any operator $\hat{A}$ acting on the qubit Hilbert space $\mathbb{C}^{2}$ as
\begin{equation}
W_{\hat{A}}(\theta,\phi))=\tr\left[\hat{A}\hat{\Delta}_{q}(\theta,\phi) \right],  
\end{equation} 
which in turn can be inverted according to
\begin{equation}\label{inverseWeyl}
\hat{A}=\frac{1}{2\pi}\int_{0}^{2\pi}\int_{0}^{\pi}W_{\hat{A}}(\theta,\phi)\hat{\Delta}_{q}(\theta,\phi)\sin\theta d\theta\,d\phi.
\end{equation}
In particular, the Wigner function for a general qubit state $\hat{\rho}=\ket{q}\bra{q}$, where
\begin{equation}
\ket{q}=a\ket{0}+b\ket{1},\;\; a,b\in\mathbb{C}, \; |a^{2}|+|b^{2}|=1,  
\end{equation} 
attains the following form
\begin{equation}
W_{\hat{\rho}}(\theta,\phi)=\tr\left[\hat{\rho}\hat{\Delta}_{q}(\theta,\phi) \right].
\end{equation}
The previous formula can alternatively be generalized for composite multi-qubit systems states \cite{Tilma}, in the context of two qubits it is formulated as 
\begin{equation}\label{Wigner2q}
	W_{\hat{\rho}}(\vec{\theta},\vec{\phi})=tr[\hat{\rho}\hat{\Delta}_{2q}(\vec{\theta},\vec{\phi})], 
\end{equation}
where $\vec{\theta}\equiv \left\lbrace \theta_{1},\theta_{2} \right\rbrace$, $\vec{\phi}\equiv \left\lbrace \phi_{1},\phi_{2} \right\rbrace$, furthermore the kernel is defined by the tensor product $\hat{\Delta}_{2q}(\vec{\theta},\vec{\phi})=\hat{\Delta}_{q}(\theta_1,\phi_1)\otimes\hat{\Delta}_{q}(\theta_2,\phi_2)$ and $\hat{\rho}$ stands for the density matrix associated to a two qubit quantum state $\ket{\psi}=a\ket{00}+b\ket{01}+c\ket{10}+d\ket{11}$, where $a,b,c,d\in\mathbb{C}$ and satisfy the normalization condition $|a|^{2}+|b|^{2}+|c|^{2}+|d|^{2}=1$. 

\noindent Finally, to close this section, it is important to point out that a noncommutative structure on the algebra of observables arises naturally from the developed phase space quantization formalism. Let $\hat{A}$ and $\hat{B}$ two operators acting on the qubit Hilbert space $\mathbb{C}^{2}$, by means of the Weyl-Stratonovich kernel (\ref{Delta}), and the equation (\ref{inverseWeyl}), the symbol associated to the product of operators $W_{\hat{A}\hat{B}}(\theta,\phi)$, determines the star product for a one qubit quantum system written as a convolution integral as \cite{Brif},
\begin{eqnarray}
\hspace{-3em}W_{\hat{A}\hat{B}}(\theta,\phi)=W_{\hat{A}}(\theta,\phi)\star W_{\hat{B}}(\theta,\phi)&=&\int_{\Omega}d\Omega'\,d\Omega''\,\,W_{\hat{A}}(\theta',\phi')W_{\hat{B}}(\theta'',\phi'') \nonumber \\
&&\times \tr\left[\hat{\Delta}_{q}(\theta,\phi)\hat{\Delta}_{q}(\theta',\phi')\hat{\Delta}_{q}(\theta'',\phi'') \right], 
\end{eqnarray}   
where $\Omega=\mathbb{S}^{2}\times\mathbb{S}^{2}$ is the region given by $0\leq \theta',\theta''\leq \pi$ and $0\leq \phi',\phi''\leq 2\pi$, and we have employed the differential $d\Omega=(1/2\pi)\sin\theta\,d\theta\,d\phi$, to denote the normalized Haar measure over a qubit phase space given by a 2-sphere. In a similar manner, this can also be generalized readily to multi-qubit systems by employing the appropriate Weyl-Stratonovich kernel. Within this formulation, the quantum dynamics of any system or composite systems can be described as
\begin{equation}
\frac{\partial W_{\hat{\rho}}}{\partial t}=\frac{1}{i\hbar}\left(W_{\hat{\rho}}\star W_{\hat{H}}-W_{\hat{H}}\star W_{\hat{\rho}} \right), 
\end{equation}
where $W_{\hat{H}}$ corresponds to the symbol of the Hamiltonian operator. This formula, known as the Moyal's equation of motion for the Wigner function, proves to be analogous to the von Neumann equation of quantum mechanics. The solution of the Moyal's equation is obtained by using the properties of the star product, it encodes all the information related to correlations functions, transition amplitudes and the evolution of the quantum dynamical system. In order to address more details on this construction we refer the reader to \cite{Zachos}, \cite{Curtright} and references therein for a detailed exposition.
  
\section{QET protocol in Phase Space Quantum Mechanics}\label{sec:QET}

In this section, we analyse the protocol called Quantum Energy Teleportation (abbreviated as QET) within the phase space formalism of quantum mechanics. The QET model, were first propose by Hotta \cite{Hotta0}, as a method to transfer energy through local operations and classical communication (LOCC) with an entangled partner, but without requiring any physical  transport between the source and the receiver. This approach maintains local energy conservation and preserves causality \cite{Hotta3},\cite{Hotta1},\cite{Hotta2},. The protocol of QET was originally developed within the framework of quantum field theory and many body systems as a technique to generate locally regions with negative energy density \cite{Hotta2}, although it has also found applications in spin chains systems \cite{Hotta0}, entangled qubits \cite{Hotta3}, on maximizing negative energy distributions for stress-energy tensor \cite{MartinEn}, algorithmic cooling protocols \cite{Cooling} and extracting energy from black holes \cite{HottaBH}, to name a few examples. Recently it has been experimentally confirmed using magnetic nuclear resonance on a bipartite quantum system \cite{Experiment1}, and in a superconducting quantum hardware \cite{Experiment2}.   

\noindent In the minimal QET scenario introduced in \cite{Hotta3}, Bob is allowed to extract some energy from a subsystem even when it starts from a passive state, such as a thermal state or the ground state. Customarily, when Bob applies a unitary local operator on its subsystem, by energy conservation, infuses energy instead of extracting it, this in fact characterizes what is known as a passive state. However, if Bob's subsystem is entangled with another subsystem shared by Alice, by performing a local measurement on Alice's states, information about quantum fluctuations of the ground state can be determined.  Next, the measurement result is communicated to Bob at a speed much faster than the diffusion velocity of the energy injected by Alice due to her measurement. Then, Bob applies conditional operators based on the announced data. Such strategy enables Bob to effectively extract energy via the ground state entanglement. It should be emphasize that the amount of output energy extracted by Bob is found to be bounded by the amount of energy infused by Alice. 

\noindent In order to analyse the minimal QET model within the phase space formulation of quantum mechanics and thus apply the techniques developed in deformation quantization, let us consider a system represented by two qubits $A$ and $B$ in the presence of a transverse magnetic field described by the Hamiltonian $\hat{H}=\hat{H}_{A}+\hat{H}_{B}+\hat{V}$, where each contribution is of the form     

\begin{equation}\label{HamiltonianoA} 
	\hat{H}_{A}=h\hat{\sigma}^{A}_{z}+\frac{h^2}{\sqrt{h^2+k^2}}\,,
\end{equation}
\begin{equation}\label{HamiltonianoB} 
	\hat{H}_{B}=h\hat{\sigma}^{B}_{z}+\frac{h^2}{\sqrt{h^2+k^2}}\,.
\end{equation}
\begin{equation}\label{HamiltonianoV}
	\hat{V}=2k\hat{\sigma}^{A}_{x}\hat{\sigma}^{B}_{x}+\frac{2k^2}{\sqrt{h^2+k^2}}\,,
\end{equation}
where $h$ and $k$ are positive constants with dimensions of energy and the operators $\sigma^{A}_z$, $\sigma^{B}_z$, $\sigma^{A}_x$ and $\sigma^{B}_x$ denote the $z$ and $x$ components of the Pauli operators for Alice and Bob respectively. Since the Hamiltonian operator is a positive-semidefinite operator, i.e. $\hat{H}\geq 0$, this implies that all its eigenvalues are non-negative \cite{Reed}. In particular, the ground state is given by 

\begin{equation}\label{ground state}
\ket{g}=\frac{1}{\sqrt{2}}\sqrt{1-\frac{h}{\sqrt{h^2+k^2}}}\ket{0}_{A}\ket{0}_{B}-\frac{1}{\sqrt{2}}\sqrt{1+\frac{h}{\sqrt{h^2+k^2}}}\ket{1}_{A}\ket{1}_{B}\,.,
\end{equation}
where $\ket{0}_{A}$, $\ket{0}_{B}$ denote the eigenstates of $\sigma^{A}_{z}$, $\sigma^{B}_{z}$ with eigenvalue $+1$ and $\ket{1}_{A}$, $\ket{1}_{B}$ stand for the eigenstates with eigenvalue $-1$ respectively. By using expression (\ref{Wigner2q}) the Wigner function associated to the ground state reads
\begin{eqnarray}\label{Wigner function base} 
\hspace{-4em}W_{\hat{\rho}_{g}}(\vec{\theta},\vec{\phi})=tr[\hat{\rho}_{g}\hat{\Delta}_{2q}(\vec{\theta},\vec{\phi})], \\
\hspace{-4em}=\frac{-\sqrt{3}}{4\sqrt{h^2+k^2}}\left( \sqrt{3}\cos(\phi_{1}+\phi_{2})\sin{\theta_1}\sin{\theta_2}+(\cos{\theta_1}+\cos\theta_2)\right) +\frac{1}{4}(3\cos{\theta_1}\cos{\theta_2}+1), \nonumber
\end{eqnarray}
where $\hat{\rho}_{g}=\ket{g}\bra{g}$ is the density operator associated to the ground state (\ref{ground state}). From the Wigner function $W_{\hat{\rho}_{g}}(\vec{\theta},\vec{\phi})$ obtained above, we can observe that it cannot be factored into two separate functions, each depending on $(\theta_{1},\phi_{1})$ or $(\theta_{2},\phi_{2})$ only. This property shows that the non-factorizability of the ground state is associated with the non-factorizability of the corresponding Wigner function. Moreover, the Wigner function depends on the variables $\phi_{1}$ an $\phi_{2}$ only through the non-classical interference term $\cos(\phi_{1}+\phi_{2})$, characteristic feature of entanglement within the phase space description \cite{Negativity}, which are related to the non-trivial topological properties of the phase space $\mathbb{S}^{2}\times\mathbb{S}^{2}$. Moreover, the Wigner function (\ref{Wigner function base}) satisfies the normalization condition
\begin{equation}
\int_{\Omega}W_{\hat{\rho}_{g}}(\vec{\theta},\vec{\phi})d\Omega_{1}\,d\Omega_{2}=1,
\end{equation} 
where $\Omega=\mathbb{S}^{2}\times\mathbb{S}^{2}$, with $0\leq \theta_{1},\theta_{2}\leq \pi$ and $0\leq \phi_{1},\phi_{2}\leq 2\pi$, and the integral Haar measure is written as $d\Omega=(1/2\pi)\sin\theta\,d\theta\,d\phi$.  
The expectation value of the Hamiltonian on the ground state can be obtained as
\begin{equation}
\braket{\hat{H}}=\braket{g,\hat{H}g}=\int_{\Omega}W_{\hat{\rho}_{g}}(\vec{\theta},\vec{\phi})W_{\hat{H}}(\vec{\theta},\vec{\phi})d\Omega_{1}\,d\Omega_{2}=0,
\end{equation}
where $W_{\hat{H}}(\vec{\theta},\vec{\phi})$ corresponds to the phase space symbol associated to the Hamiltonian operator   $\hat{H}$, and has de form
\begin{equation}\label{Weyl Hamiltonian function} 
\hspace{-3em}W_{\hat{H}}(\vec{\theta},\vec{\phi})=\sqrt{3}h(\cos\theta_{1}+\cos\theta_{2})+6k\sin\theta_{1}\sin\theta_{2}\cos\phi_{1}\cos\phi_{2}+2\sqrt{k^2+h^2}.
\end{equation}
From this expression we can observe that the first term encompasses the information of the free Hamiltonians $\hat{H}_{A}$ and $\hat{H}_{B}$, the second term incorporates the interaction $\hat{V}$, and the constant last term is added to reduce to zero the expectation value of the Hamiltonian operator for the ground state. Furthermore, the Wigner function of the ground state satisfies the star-genvalue equation
\begin{equation}
W_{\hat{H}}(\vec{\theta},\vec{\phi})\star W_{\hat{\rho}_{g}}(\vec{\theta},\vec{\phi})=0,
\end{equation} 
where the star product for a two qubit system, in the integral representation reads
\begin{eqnarray}
W_{\hat{A}}(\vec{\theta},\vec{\phi})\star W_{\hat{B}}(\vec{\theta},\vec{\phi})&=&\int_{\Omega}d\Omega_{1}'\,d\Omega_{2}'\,d\Omega_{1}''\,d\Omega_{2}''\,W_{\hat{A}}(\vec{\theta'},\vec{\phi'})W_{\hat{B}}(\vec{\theta''},\vec{\phi''}) \nonumber \\
&&\times \tr\left[\hat{\Delta}_{2q}(\vec{\theta},\vec{\phi})\hat{\Delta}_{2q}(\vec{\theta'},\vec{\phi'})\hat{\Delta}_{2q}(\vec{\theta''},\vec{\phi''}) \right], 
\end{eqnarray}  
where the region $\Omega=(\mathbb{S}^{2}\times\mathbb{S}^{2})^{2}$ and $d\Omega$ represents the normalized Haar measure of a qubit phase space as state previously.

\noindent Following the QET protocol, Alice performs a projective measurement of the observable $\hat\sigma_{x}^{A}$ on the ground state with measurement output $\alpha=\pm 1$. The projection operator corresponding to each measurement outcome has the form
\begin{equation}
\hat{P}^{A}(\alpha)=\frac{1}{2}\left(\hat{1}+\alpha\hat{\sigma}_{x}^{A} \right), 
\end{equation}
and its associated phase space symbol reads
\begin{equation}\label{WignerProjection}
W_{\hat{P}^{A}(\alpha)}(\vec{\theta},\vec{\phi})=\frac{1}{2}\left(1+\alpha\sqrt{3}\sin\theta_{1}\cos\phi_{1}\right). 
\end{equation}
After the measurement, the state of the two qubits is transformed into the density operator
\begin{equation}\label{Density step1} 
\hat{\rho}'=\sum_{\alpha=\pm 1} \hat{P}^{A}({\alpha})\hat{\rho}_{g}\hat{P}^{A}({\alpha}), 
\end{equation}
which, within the phase space formalism, demonstrates to be equivalent to
\begin{eqnarray}\label{Wigner function step1} 
\hspace{-3em}W_{\hat{\rho}'}(\vec{\theta},\vec{\phi})&=&\sum_{\alpha=\pm 1}W_{\hat{P}^{A}(\alpha)}(\vec{\theta},\vec{\phi})\star W_{\hat{\rho}_{g}}(\vec{\theta},\vec{\phi})\star W_{\hat{P}^{A}(\alpha)}(\vec{\theta},\vec{\phi}), \nonumber \\
&=&-\frac{\sqrt{3}h}{4\sqrt{h^2+k^2}}\cos\theta_{2}-\frac{3k}{4\sqrt{h^2+k^2}}\sin\theta_{1}\sin\theta_{2}\cos\phi_{1}\cos\phi_{2}+\frac{1}{4}.
\end{eqnarray}
Since the resultant Wigner function no longer describes the ground state, a certain amount of energy has been infused by Alice into the system. This energy can be obtained as
\begin{equation}
E_{A}=\int_{\Omega}W_{\hat{\rho}'}(\vec{\theta},\vec{\phi})W_{\hat{H}}(\vec{\theta},\vec{\phi})d\Omega_{1}\,d\Omega_{2}=  \frac{h^2}{\sqrt{h^2+k^2}}.
\end{equation}
The energy $E_{A}$ is regarded as the QET energy input resulting from the Alice measurement \cite{Hotta3}, and as we will discuss later, this energy proves to be proportional to the amount of entanglement in the initial state. The key aspect of this protocol stems form the fact that any measurement performed by Alice on her subsystem does not increase the average energy of the Bob's subsystem, i.e., the average values of the operators $\hat{H}_{B}$ and $\hat{V}$ equate to zero, since
\begin{equation}
\braket{\hat{H}_{B}}_{\hat{\rho}'}=\int_{\Omega}W_{\hat{\rho}'_{g}}(\vec{\theta},\vec{\phi})W_{\hat{H}_{B}}(\vec{\theta},\vec{\phi})d\Omega_{1}\,d\Omega_{2}=0,
\end{equation} 
and
\begin{equation}
\braket{\hat{V}}_{\hat{\rho}'}=\int_{\Omega}W_{\hat{\rho}'_{g}}(\vec{\theta},\vec{\phi})W_{\hat{V}}(\vec{\theta},\vec{\phi})d\Omega_{1}\,d\Omega_{2}=0.
\end{equation} 
Consequently, there are no instantaneous force acting from the subsystem A to the subsystem B after the measurement. Because the model is non-relativistic, the protocol allows Bob to extract energy faster than the diffusion speed of the input energy given by Alice measurement, which proves to be proportional to $k$ \cite{Hotta0}. Next, according to the QET  protocol, after Alice's measurement result is communicated to Bob, Bob applies a local conditional unitary operator given by
\begin{equation}
\hat{U}^{B}(\alpha)=\hat{1}_{B}\cos\omega-i\alpha\sin\omega\hat{\sigma}^{B}_{y},
\end{equation}  
where $\alpha=\pm 1$ and $\omega$ is a real constant that satisfies
\begin{equation}
\cos\omega=\frac{h^{2}+2k^{2}}{\sqrt{(h^{2}+2k^{2})^{2}+h^{2}k^{2}}},
\end{equation}
and
\begin{equation}
\sin\omega=\frac{hk}{\sqrt{(h^{2}+2k^{2})^{2}+h^{2}k^{2}}}.
\end{equation}
The local unitary operator $\hat{U}^{B}(\alpha)$ is represented by the associated Weyl symbol on the phase space as
\begin{equation}
W_{\hat{U}^{B}(\alpha)}(\vec{\theta},\vec{\phi})=\cos\omega-i\sqrt{3}\alpha\sin\omega\sin\theta_{2}\sin\phi_{2}.
\end{equation}
Then, after the application of the $\alpha$-dependent local unitary operator, the Wigner function of the system reads
\begin{eqnarray}\label{Wigner function step2} 
\hspace{-4em}W_{\hat{\rho}''}(\vec{\theta},\vec{\phi})&=&\sum_{\alpha=\pm 1}W_{\hat{U}^{A}(\alpha)}(\vec{\theta},\vec{\phi})\star W_{\hat{\rho}'}(\vec{\theta},\vec{\phi})\star W^{*}_{\hat{U}^{A}(\alpha)}(\vec{\theta},\vec{\phi}), \nonumber \\
&=&-\frac{\sqrt{3}h}{4\sqrt{h^2+4k^2}}\cos\theta_{2}-\frac{3k}{2\sqrt{h^2+4k^2}}\sin\theta_{1}\sin\theta_{2}\cos\phi_{1}\cos\phi_{2}+\frac{1}{4}.
\end{eqnarray}
Finally, the output energy extracted by Bob using the QET protocol can be computed as follows
\begin{eqnarray}
E_{B}&=& \braket{\hat{H}}_{\hat{\rho}''}-E_{A}=\int_{\Omega}W_{\hat{\rho}''}(\vec{\theta},\vec{\phi})W_{\hat{H}}(\vec{\theta},\vec{\phi})d\Omega_{1}\,d\Omega_{2}-E_{A},\nonumber \\
&=& \frac{h^2+2k^2}{\sqrt{h^2+k^2}}\left[ \sqrt{1+\frac{h^2k^2}{(h^2+2k^2)^2}}-1\right] 
\end{eqnarray}
which is evidently non-negative. It is noteworthy to observe that the teleported energy proves to be proportional to the amount of entanglement given in the initial ground state. To illustrate this, let us consider the negativity as an entanglement measure of a system. It can be defined in terms of the density matrix as a monotone positive functional defined by
\begin{equation}
\mathcal{N}(\hat{\rho})=\frac{||\hat{\rho}^{t_{B}}||-1}{2},
\end{equation}     
where $\hat{\rho}^{t_{B}}$ denotes the partial transpose of the density operator with respect to the system $B$, and $||\,.\,||$ denotes the trace norm \cite{Nielsen}. The importance of this measure arises from the fact that the logarithm of the trace norm of the partial transpose serves as an upper bound for distillable entanglement \cite{Horodecki}. It is straightforward to check  that the negativity of the ground state reads
\begin{equation}
\mathcal{N}(\hat{\rho}_{g})=\frac{k}{2\sqrt{k^2+h^{2}}},
\end{equation} 
meanwhile for the final step, the negativity is zero, i.e., $\mathcal{N}(\hat{\rho}'')=0$. Hence, the teleported energy can be written as
\begin{equation}
E_{B}=\frac{2\mathcal{N}(\hat{\rho}_{g})}{k}\left[\sqrt{h^{2}k^{2}+(h^{2}+2k^{2})^2}-(h^{2}+2k^{2})\right],
\end{equation}
this means that in order to implement the QET protocol, it is required to consume some amount of the ground state entanglement. An alternative way to understand this aspect is using the so called Wehrl entropy in terms of the $Q$-function \cite{Wehrl}, \cite{Obada}. In a two qubit system, let us consider a Bloch coherent state defined as
\begin{eqnarray}
\ket{\Phi}=&&e^{i(\phi_{1}+\phi_{2})}\sin\frac{\theta_{1}}{2}\sin\frac{\theta_{2}}{2}\ket{0}_{A}\ket{0}_{B}+e^{i\phi_{1}}\sin\frac{\theta_{1}}{2}\cos\frac{\theta_{2}}{2}\ket{0}_{A}\ket{1}_{B} \nonumber \\
&&+e^{i\phi_{2}}\cos\frac{\theta_{1}}{2}\sin\frac{\theta_{2}}{2}\ket{1}_{A}\ket{0}_{B}+\cos\frac{\theta_{1}}{2}\cos\frac{\theta_{2}}{2}\ket{1}_{A}\ket{1}_{B},
\end{eqnarray}
where $0\leq \theta_{1},\theta_{2}\leq \pi$ and $0\leq \phi_{1},\phi_{2}\leq 2\pi$. The Husimi quasi-probability distribution, also known as the $Q$-function, consists of a non-negative function defined on the phase space which, analogously to the Wigner function, enables a complete characterization of a quantum system. The $Q$-function is primarily used in quantum optics and tomography to analyse different quantum effects in the coupling of light-matter interaction and superconductors \cite{Zachos}, \cite{Husimi}. For a two qubit system, it is given by
\begin{eqnarray}
Q_{\hat{\rho}}(\vec{\theta},\vec{\phi})&=&\braket{\Phi,\hat{\rho}\Phi}, \nonumber \\
&=&\int_{\Omega}W_{\hat{\rho}}(\vec{\theta},\vec{\phi})\braket{\Phi,\hat{\Delta}_{2q}(\vec{\theta},\vec{\phi})\Phi}d\Omega_{1}\,d\Omega_{2},
\end{eqnarray}
whereas the partial $Q$-function of the B qubit is expressed as
\begin{equation}
Q^{B}_{\hat{\rho}}(\theta_{2},\phi_{2})=\int_{{\mathbb{S}^{2}}}Q_{\hat{\rho}}(\vec{\theta},\vec{\phi})\,d\Omega_{1}.
\end{equation}  
On the other hand, the Wehrl entropy was introduced in \cite{Wehrl} as a measure of diverse quantum properties, including quantum noise \cite{Buzek}, decoherence and entanglement \cite{BuzekII}. The partial Wehrl entropy of the qubit B is written in terms of the $Q$-function as
\begin{equation}
E^{B}_{\hat{\rho}}=\int_{\mathbb{S}^{2}}-Q^{B}_{\hat{\rho}}(\theta_{2},\phi_{2})\ln Q^{B}_{\hat{\rho}}(\theta_{2},\phi_{2})\,d\Omega_{2}. 
\end{equation}    
Here the Wehrl entropy is used as a useful measure of the loss of entanglement of a state as a consequence of the unitarities applied during the protocol. We can notice that the Wehrl entropy cannot be negative and exhibits a temporal evolution similar to the von Neumann entropy \cite{Obada}. For the initially prepared ground state, the Wehrl entropy satisfies
\begin{equation}
2.3379=\frac{1}{4\pi}(1+\ln 4+2\ln\pi)\leq E_{\hat{\rho}_{g}}^{B}\leq\ln 4\pi,
\end{equation}
where $E_{\hat{\rho}_{g}}^{B}$ takes the form
\begin{equation}
\hspace{-4em}E_{\hat{\rho}_{g}}^{B}=\frac{2h^{2}+k^{2}}{2\pi hk}\mathcal{N}(\hat{\rho}_{g})\arctan\left( \frac{2h}{k}\mathcal{N}(\hat{\rho}_{g})\right)-\frac{1}{4\pi}\left(1+\ln\left(16+\frac{16 h^{2}}{k^{2}}\right)+2\ln\pi  \right).  
\end{equation}
The minimal value of the Wehrl entropy for the ground state occurs at $k=0$, i.e., on a separable pure state, while the maximum value takes place at $h=0$, which represents a maximally entangled initial state. Moreover, the entropy associated to the final state of the protocol proves to be $ E_{\hat{\rho}''}^{B}=\ln 4\pi$. Since the entropy of the system cannot decrease, it follows that 
\begin{equation}
2.3379\leq  E_{\hat{\rho}_{g}}^{B}\leq  E_{\hat{\rho}'}^{B}\leq  E_{\hat{\rho}''}^{B}=\ln 4\pi.
\end{equation} 
Considering that the negativity of the final state is equal to zero, this implies that the Wehrl entropy quantifies the consumption of coherence and entanglement along the protocol for the purpose of extracting energy. Furthermore, the amount of entropy at the final step is fully described by classical correlations, which were originated from the conditional operators applied by Bob, based on Alice measurements, via a classical communication channel.  

\section{Conclusions}
\label{sec:conclu}

In this paper we have analysed the QET protocol within the phase space formulation of quantum mechanics. Following some ideas introduced by the Weyl-Wigner quantization of finite dimensional spin systems, we have obtained the Wigner function and the star-product of the minimal QET model, which corresponds to a pair of two entangled qubits in the presence of a transverse magnetic field. Then, by using the integral properties of the Winger quasi-probability distribution and the symbols corresponding to a sequence of unitarities, we showed that the teleported energy is proportional to the amount of entanglement present in the initial ground state, which is consistent with previous results. Finally, with the aid of the Husimi $Q$-function, we analysed the Wehrl entropy of the protocol. As we have seen, the Wehrl entropy enables us to quantify the consumption of coherence and entanglement of the initial ground state in order to extract energy, resulting in a final state with purely classical correlations. We expect the results established here may clarify several aspects of the energy teleportation protocols via local operations and classical communication, found in the literature. As future work, we pretend to implement our quantization approach to the case of quantum fields, in order to explore how quasi-probability distributions describe the violation of classical energy conditions. This will enable us to investigate the QET protocol in more general scenarios. This will be done elsewhere.

\section*{Acknowledgments}
The authors would like to acknowledge support from CONAHCYT-Mexico. JBM thanks to the Dipartimento di Fisica ``Ettore
Pancini" for the kind invitation and its generous hospitality. Computations have been made using wxMaxima.

\section*{References}

\bibliographystyle{unsrt}

\end{document}